\documentclass[twocolumn,aps,prc,superscriptaddress,showpacs]{revtex4}
\usepackage{amsmath,bm}%
\usepackage{graphicx}%

\begin{document}

\draft
\title{  Effect of Hadronic Rescattering on Elliptic Flow  Following Hydrodynamics Model}

\author {G. L. Ma}
\affiliation{Shanghai Institute of Applied Physics, Chinese
Academy of Sciences, P.O. Box 800-204, Shanghai 201800, China}
\affiliation{Graduate School of the Chinese Academy of Sciences,
Beijing 100080, China}
\author{Y. G. Ma}
\thanks{Corresponding author: ygma@sinap.ac.cn}
\affiliation{Shanghai Institute of Applied Physics, Chinese
Academy of Sciences, P.O. Box 800-204, Shanghai 201800, China}
\author{B. H. Sa}
\affiliation{China Institute of Atomic Energy, P.O. Box 918,
Beijing 102413, China}
\author{X. Z. Cai}
\author{J. H. Chen}
\affiliation{Shanghai Institute of Applied Physics, Chinese
Academy of Sciences, P.O. Box 800-204, Shanghai 201800,
China}
\author{Z. J. He} \affiliation{Shanghai Institute of Applied
Physics, Chinese Academy of Sciences, P.O. Box 800-204, Shanghai
201800, China}
\author{J. L. Long}
\author{W. Q. Shen}
\author{C. Zhong}
\author{J. X. Zuo}
\author{J. G. Chen}
\author{D. Q. Fang}
\author{W. Guo}
\author{C. W. Ma}
\author{Q. M. Su}
\author{W. D. Tian}
\author{K. Wang}
\author{Y. B. Wei}
\affiliation{Shanghai Institute of Applied Physics, Chinese
Academy of Sciences, P.O. Box 800-204, Shanghai 201800, China}


\date{\today}

\begin{abstract}
The effect of hadronic rescattering on the elliptic flow has been
investigated by the Cooper-Frye hadronization model from
hydrodynamic evolution following by the afterburner hadronic
rescattering model for 200 GeV/c Au + Au at 20-40$\%$ centrality.
It is found that the hadronic rescattering can suppress elliptic
flow $v_2$ in momentum space especially in lower transverse
momentum region.   In addition, the hadronic rescattering effects
on transverse momentum spectra and anisotropy coordinate space of
hadrons are studied.
\end{abstract}



\maketitle

\section{ Introduction}

Lattice QCD calculations have predicted a transition from
quark-gluon matter (QGP) to hadronic matter at high temperature or
density \cite{Wilson,Satz}. This transition was believed to occur
around ten microseconds after Big Bang. Practically, one may
obtain this piece of information by studying the relativistic
heavy ion collisions at Brookhaven National Laboratory now. Some
possible probes have been proposed for searching for it
\cite{Bass,Reit}. Elliptic flow $v_2$ is one of these probes, from
which one may understand some early information in relativistic
heavy ion collision through measuring anisotropy of momentum space
of particles in final state. Some experimental results from
BNL-RHIC have been published    recently
\cite{Adler1,Acker,Alder1,Alder2,Alder3,Adcox,Adams}.

As we know, hydrodynamical model, which is one method to study
relativistic heavy ion collisions, can reproduce  $v_2$ data very
well below $\sim$ 2 GeV/c of transverse momentum ($p_T$)
\cite{Kolb1,Kolb2,Huovinen,Csanad,Hirano}. However, the hadronical
rescattering effect on elliptic flow after the hadronization
following by the hydrodynamics, at least in our knowledge, has not
been investigated in details so far. Based on this motivation,  we
shall focus the influence of rescattering in hadronic state on
elliptic flow $v_2$. It should be pointed
    that hadronic phase space before rescattering is produced directly by an anisotropic
     momentum space of produced hadrons in order to produce $v_2$ in our calculation,
     which will be described in detail in following. As a result, we find that $v_2$ of final
     particles is suppressed due to the rescattering among hadrons to a considerable
     extent. It may indicate that the effect of rescattering among hadrons on $v_2$ is not
     ignored in the research on $v_2$.

\section{ Hydrodynamics model and Initial phase space of hadrons}

Hydrodynamics of relativistic heavy ion collision
\cite{Rischke1,Rischke2,Dumitru1}
     is defined by (local) energy-momentum and net charge conservation,
\begin{equation}\label{1}
\partial_{\mu }T^{\mu\nu} = 0,\quad   \partial_{\mu }N_{i}^{\mu} = 0,
\end{equation}
where $T^{\mu\nu}$ denotes the energy-momentum tensor and
$N_{i}^{\mu}$ the net four-current of the $i$th conserved charge.
Here will only take the net baryon number. We implicitly assume
that all other charges which are conserved on strong-interaction
time scale, e.g., strangeness, charm, and electric charge, vanish
locally. The corresponding four-currents are therefore identically
zero, and the conservation equations are trivial.

  For ideal fluids, the energy-momentum tensor and net
  baryon current assume the simple form:
\begin{equation}\label{2}
T^{\mu\nu} = (\epsilon+p)u^{\mu}u^{\nu}-pg^{\mu\nu},\quad
N_{B}^{\mu} = \rho_{B}u^{\mu},
\end{equation}
where $\epsilon$, p, and $\rho_{B}$ are energy density, pressure,
and net baryon density in the local rest frame of the fluid, which
is defined by $N_{B}^{\mu}$ = ($\rho_{B}$,$\overrightarrow{0}$),
respectively. Here $g^{\mu\nu}$ = diag(+,-,-,-) is the metric
tensor, and $u^{\mu}$ = $\gamma$(1,$\overrightarrow{v}$) the
four-velocity of the fluid ($\overrightarrow{v}$ is the
three-velocity and $\gamma$ = $(1-\overrightarrow{v}^{2})^{-1/2}$,
the Lorentz factor).

For simplicity, the hydrodynamics that we are using  assumes
cylindrically symmetric transverse expansion with a longitudinal
scaling flow profile, $v_z = z/t$ though we are investigating the
non-central collision. We recognize that  our initial phase space
of hadrons in this way is not perfect since we put an initial
anisotropy of phase space by hand  just after Cooper-Frye
hadronization as illustrated in the following. Nevertheless, we
think that it is still useful and interesting to investigate the
hadronic rescattering effect on spectra and flow based on this
treatment. In this point, our simulation is somehow like that we
start from an assumed initial phase space of hadrons and then
study the hadronic rescattering effect. This phase space is
similar to a sudden hadronization phase space which the partonic
anisotropy turns to hadron anisotropy. For hydrodynamics, at $z$ =
0, Eqs (1) reduces to
\begin{equation}
\partial_{t}E+\partial_{T}[(E+p)v_{T}] = -(\frac{v_{T}}{r_{T}}+\frac{1}{t})(E+p),
\end{equation}
\begin{equation}
\partial_{t}M+\partial_{T}[M v_{T}+p] = -(\frac{v_{T}}{r_{T}}+\frac{1}{t})M,
\end{equation}
\begin{equation}
\partial_{t}R+\partial_{T}[R v_{T}] = -(\frac{v_{T}}{r_{T}}+\frac{1}{t})R,
\end{equation}
where we defined E $\equiv$ $T^{00}$, M $\equiv$ $T^{0T}$, and R
$\equiv$ $N_{B}^{0}$. In the above expressions, the index $T$
refers to the transverse component of the corresponding quantity.
The equations (3)-(5) describe the evolution in the $z$ = 0 plane.
Because of the assumption of longitudinal scaling, the solution at
any other $z \neq$ 0 can be simply obtained by a Lorentz boost.

   Initial conditions for Au + Au at 200 GeV/c in 20-40$\%$ centrality  is
supposed that initial  energy density  = 10.3 $GeV/fm^3$ and
initial time   = 0.6 $fm/c$, and a hydrodynamical
    evolution of cylindrically symmetric transverse expansion with a longitudinal
     scaling flow profile is adopted. When temperature of liquid reaches to a
     critical energy density ( = 0.42 $GeV/fm^3$), a hadronization of QGP which is described
     by employing the MIT bag  model EOS \cite{Chodos} will take place. And an ideal hadron gas
     is established and the corresponding EOS is applied. Hadronization in hydrodynamics
     usually takes the method of Cooper and Frye \cite{Cooper}.
     In this work, however, we make an assumption that the coordinate and momentum spaces of
     hadronized particles are not isotropic but have shapes as plotted in
      Figure 1 in order to simulate Au + Au at 200 GeV/c in  20-40$\%$ centrality and get $v_2$ in our
       calculation. Here the mean anisotropy in momentum space, $F \equiv <$Px/Py$>$, takes 1.18 in
       our simulation for all produced hadrons.

For the details of the simulation to result in the above
anisotropic distributions, we wrote
\begin{equation*}
R_{x}=R_{Tx}cos\theta  ,\quad   P_{x}=P_{Tx}\cos\psi
\end{equation*}
\begin{equation*}
R_{y}=R_{Ty}sin\theta          ,\quad   P_{y}=P_{Ty}\sin\psi
\end{equation*}
\begin{equation*}
R_{z}=t *tanh y           , \quad  P_{z}=m_{T}\sinh y
\end{equation*}

\begin{equation}
t=\xi tanh y          ,\quad
E=\sqrt{P_{x}^2+P_{y}^2+P_{z}^2+m_{0}^{2}}
\end{equation}

\begin{equation}
<P_{Tx}/P_{Ty}>=<R_{Ty}/R_{Tx}>=F
\end{equation}

\begin{equation}
<\theta / \psi >=1
\end{equation}
 where $m_{T} = \sqrt{p_{T}^{2}+m_{0}^{2}}$,
 ($R_{x},R_{y},R_{z},t$) and ($P_{x},P_{y},P_{z},E$) are the
 4-vectors of  produced hadrons in coordinate and momentum spaces respectively.
 We assume that hadrons are randomly distributed in the overlap coordinate space of two
spherical nucleus in terms of the left column of equations (7) by
hadronization. $\xi$ and y are random numbers which are
respectively randomly distributed in [0,14] fm/c and [-5,5]. As
for momentum and position of hadrons, $P_{Tx}$ and $R_{Ty}$ denote
a random variable distributed in [0,4] GeV and [0,7] fm, then
$P_{Ty}$ and $R_{Tx}$ can be obtained by the anisotropic factor in
momentum and coordinate space, namely $F$, which makes our
momentum and coordinate space distribution in transverse plane
elliptic and anti-elliptic. If we make $F$ equal to 1, our
hadronization method will go back to the method of Cooper and
Frye.  $\theta$ and $\psi$ are the azimuths distributed in
[0,2$\pi$] in coordinate and momentum spaces. But here we assume
these produced hadrons are from a radial source in which $\theta$
and $\psi$ have a correlation like the equation (8), which means
that once a $\psi$ has been decided, the corresponding $\theta$
will be selected in a Gaussian distribution whose mean is  $\psi$
and whose width is always assumed to be 0.2$\pi$ in our
calculation. The set of equations results in the coordinate and
momentum space as figure 1 shows. Here we choose the anisotropic
factor $F$ in momentum space, equals to 1.18 for 200 GeV/c Au + Au
in 20-40$\%$ centrality.

\begin{figure}
\includegraphics[scale=0.45]{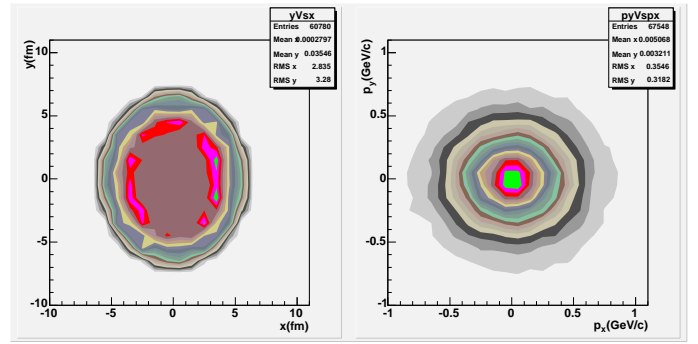}
\caption{\footnotesize Contour plots of shapes of the phase space
where hadrons are produced when two same spherical overlap in
half-central collision. Left: transverse plane at z = 0 in
coordinate space; Right: transverse plane at Pz = 0 in momentum
space.
 }
\label{shape}
\end{figure}

\section{Hadronic rescattering model}

After the hadrons are produced, they enter next  rescattering
process among them. We here use the rescattering model from
  LUCIAE model \cite{An,Sa}. Two particles will collide if
  their minimum distance $d_{min}$ satisfies
\begin{equation}\label{dmin}
d_{min} \leq  \sqrt{\frac{\sigma_{total}}{\pi}},
\end{equation}
where $\sigma_{total}$ is the total cross section in $fm^{2}$ and
the minimum distance is calculated in the C.M.S. frame of the two
colliding particles. If these two particles are moving towards
each other at the time when both of them are formed, the minimum
distance is defined as the distance perpendicular to the momentum
of both particles. If the two particles are moving back-to-back,
the minimum distance is defined as the distance at the moment when
both of them are formed. Assuming that the hadrons move along
straight-line classical trajectories between two consecutive
collisions it is possible to calculate the collision time when two
hadrons reach their minimum distance and order all the possible
collision pairs according to the collision time sequence. If the
total and the elastic cross section satisfies
\begin{equation}\label{posb}
\frac{\sigma_{elastic}}{\sigma_{total}}\geq \eta
\end{equation}
where $\eta$ is a random number then the particles will be
elastically scattered or else the collision will be considered as
an inelastic collision. The distribution of the momentum transfer,
 $t$, is taken as
\begin{equation}\label{t}
\frac{d\sigma}{dt}\sim exp(Bt)
\end{equation}
where $B$, for an elastic scattering, depends on the masses of two
scattering particles. The azimuthal angle will be isotropically
distributed.

During the rescattering process, the following inelastic reactions
are included in our calculation:
\\

  \begin{tabular}{|c|c|}
  \hline
  $\pi$N $\rightleftharpoons$ $\Delta$ $\pi$  & $\pi$N $\rightleftharpoons$ $\rho$ N \\
  N N $\rightleftharpoons$ $\Delta$ N & $\pi$ $\pi$ $\rightleftharpoons$k $\overline{k}$  \\
  $\pi$ N $\rightleftharpoons$ k Y & $\pi$ $\overline{N}$ $\rightleftharpoons$ $\overline{k}$ $\overline{Y}$ \\
  $\pi$ Y $\rightleftharpoons$ k $\Xi$ & $\pi$ $\overline{Y}$ $\rightleftharpoons$ $\overline{k}$ $\overline{\Xi}$\\
  $\overline{k}$ N $\rightleftharpoons$ $\pi$ Y & k $\overline{N}$ $\rightleftharpoons$ $\pi$ $\overline{Y}$\\
  $\overline{k}$ Y $\rightleftharpoons$ $\pi$ $\Xi$ & k $\overline{Y}$ $\rightleftharpoons$ $\pi$ $\overline{\Xi}$\\
  $\overline{k}$ N $\rightleftharpoons$ k $\Xi$ & k $\overline{N}$ $\rightleftharpoons$ $\overline{k}$ $\overline{\Xi}$\\
  $\pi$ $\Xi$ $\rightleftharpoons$ k $\Omega^{-}$ & $\pi$ $\overline{\Xi}$ $\rightleftharpoons$ $\overline{k}$ $\overline{\Omega^{-}}$\\
  k $\overline{\Xi}$ $\rightleftharpoons$ $\pi$ $\overline{\Omega^{-}}$ & $\overline{k}$ $\Xi$ $\rightleftharpoons$ $\pi$ $\Omega^{-}$\\
  $\overline{N}$ N annihilation & $\overline{Y}$ N annihilation \\
  \hline
\end{tabular}
\vspace{0.5cm}

 \hspace{-.45cm} where the hyperons $Y$ =
$\Lambda$ or $\Sigma$. The relative probabilities for the
different channels, e.g. in ($\pi$ N)-scattering, is used to
determine the outcome of the inelastic encounter. As the reactions
introduced above do not make up the full inelastic cross section,
the remainder is again treated as elastic encounters.

 \section { Results and Discussions}

 Figure 2 shows the transverse momentum distribution of
$\pi^-$, $k^{-}$, $\overline{P}$ before and after rescattering in
Au + Au at 200GeV/c in 20-40$\%$ centrality  in our hydrodynamical
calculation followed by the rescattering model.

\begin{figure}
\includegraphics[scale=0.85]{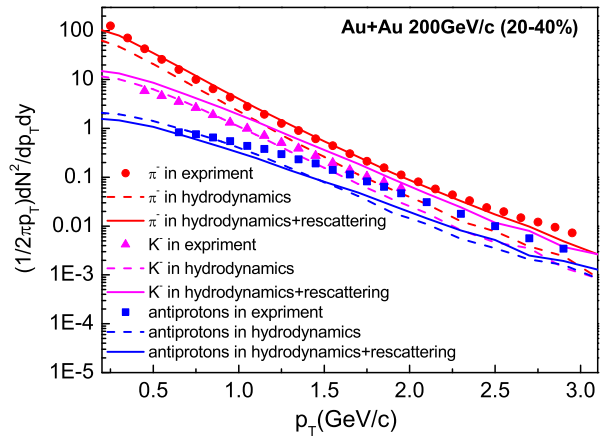}
\caption{\footnotesize The transverse momentum distributions of
$\pi^-$, $k^{-}$, $\overline{P}$  in Au + Au at 200GeV/c in
20-40$\%$ centrality in our hydrodynamical + rescattering model.
Full lines and dash lines represent $p_T$ spectra after and before
hadronic rescattering, respectively; solid symbols are
experimental data which come from \cite{Adler2}.} \label{Pt}
\end{figure}

At the same time, we obtain the $v_2$ of hadrons before and after
rescattering in terms of formula (12), which indicates the
anisotropy of hadrons' momentum space.
\begin{equation}\label{V2}
v_2 \equiv <\frac{p_{x}^{2}-p_{y}^{2}}{p_{x}^{2}+p_{y}^{2}}>
\end{equation}
Figure 3   shows dependence of $v_2$ of $\pi^-$+$k^{-}$,
$h^{-}$+$h^{+}$, $\overline{P}$, $\Lambda$+$\overline{\Lambda}$ on
$p_T$ before and after rescattering in hydrodynamics and
experimental data in Au+Au at 200GeV/c in 20-40$\%$ centrality
\cite{Adler2}.
\begin{figure}
\includegraphics[scale=0.8]{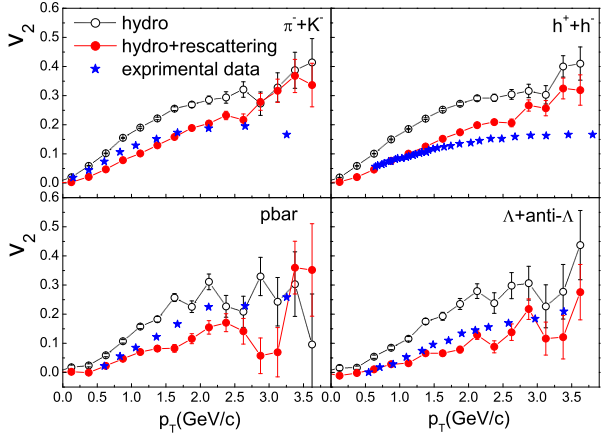}
\caption{\footnotesize The $p_T$ dependences of $v_2$ of
$\pi^-$+$k^{-}$, $h^{-}$+$h^{+}$, $\overline{P}$ and  $\Lambda$ +
$\overline{\Lambda}$ before and after rescattering in
hydrodynamical+rescattering model and experimental data which come
from \cite{Adler1} and \cite{Adams} in Au+Au at 200 GeV/c in
20-40$\%$ centrality. } \label{V2}
\end{figure}

In order to study anisotropy of hadrons' coordinate space, the
parameter named as $\epsilon_{2}$ is defined by formula (13).
\begin{equation}\label{$v_2$r}
\epsilon_{2} \equiv <\frac{x^{2}-y^{2}}{x^{2}+y^{2}}>
\end{equation}

\begin{figure}
\includegraphics[scale=1.05]{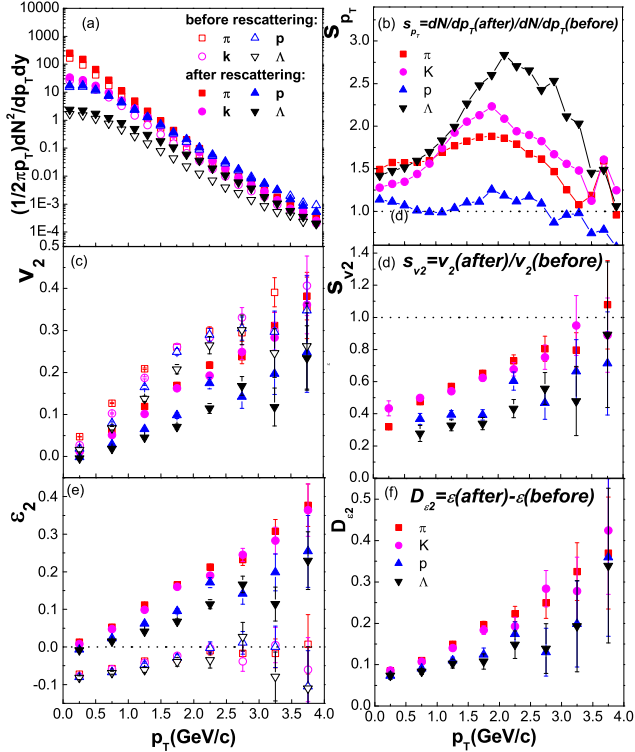}
\caption{\footnotesize (a),(c) and (e): The $p_T$ dependences of
spectra, $v_2$ and $\epsilon_{2}$ of $\pi^+$+$\pi^{-}$ (square),
$k^{+}$+$k^{-}$ (circle), P+$\overline{P}$ (up-triangle) and
$\Lambda$+$\overline{\Lambda}$ (down-triangle) before (open
symbols) and after (full symbols) rescattering in hydrodynamical +
rescattering model in Au+Au at 200 GeV/c in 20-40$\%$ centrality;
(b),(d) and(f): The  $p_T$ dependences of  $s_{P_T}$, $s_{v_2}$
and $D_{\epsilon_{2}}$. See more details in text.} \label{V2}
\end{figure}

The plots (a) (c) and (e) in figure 4 give $p_T$ spectra, $v_2$
and $\epsilon_{2}$ of four types of particles ((
$\pi^+$+$\pi^{-}$), ($k^{+}$+$k^{-}$), (P+$\overline{P}$) and
($\Lambda$+$\overline{\Lambda}$) ) vs $p_T$ before and after
rescattering in hydrodynamics+rescattering model. The open and
filled points, respectively,  stand for the status before and
after hadrons' rescattering. We can see that $p_T$ yield and
$\epsilon_{2}$ of hadrons is enhanced by rescattering, and also
note $\epsilon_{2}$ changes its sign through rescattering. On the
other hand, elliptic flow $v_2$ of hadrons is suppressed by the
rescattering. In order to quantify the extent of enhancement (or
suppression ) at different $p_T$ bins, we define three factors
named as $s_{P_T}$, $s_{v_2}$ and $D_{\epsilon_{2}}$ by formula
(14),(15) and (16).

\begin{equation}
s_{P_T}(p_T) \equiv \frac{\frac{dN}{dydP_{T}}_{after
\hspace{0.05in} }}{\frac{dN}{dydP_{T}}_{before \hspace{0.05in} }}
\end{equation}
\begin{equation}
s_{v_2}(p_T) \equiv \frac{V_2(p_T)_{after \hspace{0.05in}
}}{V_2(p_T)_{before \hspace{0.05in} }}
\end{equation}
\begin{equation}
D_{\epsilon_{2}}(p_T) \equiv
\epsilon(p_T)_{after}-\epsilon(p_T)_{before }
\end{equation}

The plots (b) (d) and (f) in figure 4 show respective dependence
of $s_{P_T}$ $s_{v_2}$ and $D_{\epsilon_{2}}$on $p_T$. In (b), we
find that the yield of hadrons is increased by hadrons'
rescattering, especially for higher $p_T$ yield. It indicates the
rescattering produced many secondary hadrons,whose $p_T$ seems
harder than those hadrons in last generation. In (d), $s_{v_2}$
increases with $p_T$, which indicates that the suppression effect
on $v_2$ from  hadrons' rescattering becomes weaker and weaker
with the increasing of $p_T$. In (e) and (f), we see the
rescattering turns the shape of coordinate space, and the effect
becomes stronger with the increasing of $p_T$. It indicates the
rescattering makes hadrons' momentum space less anisotropy and
more heavier particles, more suppression the $v_2$. While the
changes of the tropism of coordinate space occurs in the same
time. In our calculation, at least 20 $\sim$ 40\% $v_2$ will
disappear through final hadrons' rescattering. So the effect from
final hadrons' rescattering on $v_2$ maybe should be considered in
order to research early state information in relativistic heavy
ion collisions.

\begin{figure}
\includegraphics[scale=1.05]{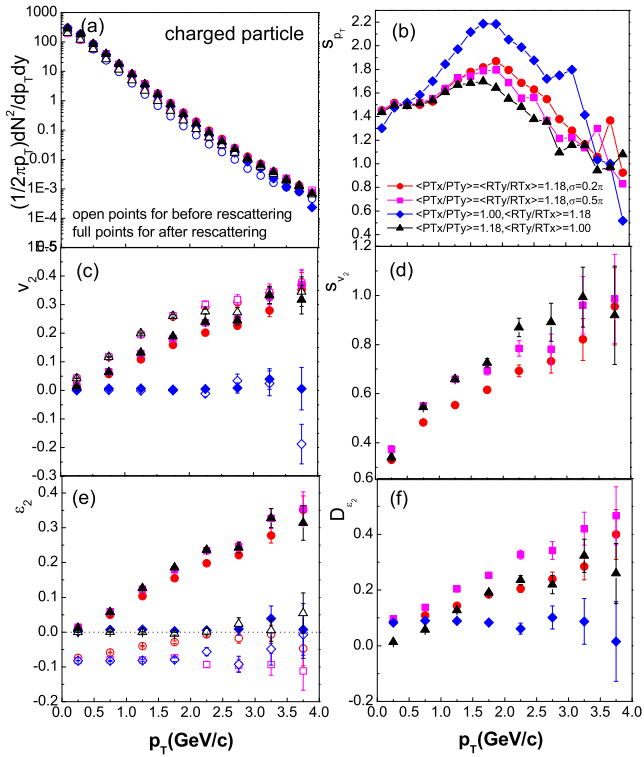}
\caption{\footnotesize (a), (c) and (e): The $p_T$ dependences of
spectra, $v_2$ and $\epsilon_{2}$ of charged particle before (open
symbols) and after (full symbols) rescattering in hydrodynamical +
rescattering model in Au + Au at 200 GeV/c in 20-40$\%$
centrality. The circles and squares are for the hadronization with
both initial $v_2$ and $\epsilon_{2}$, and the circles correspond
to 0.2$\pi$  of the width of $\psi$-$\phi$ correlation and squares
correspond to 0.5$\pi$. The diamonds  are from the hadronization
whose particles are only with initial $\epsilon_{2}$and the
triangles are from the hadronization whose particles are only with
initial $v_2$ (see the insert of (b)); (b),(d) and (f): The $p_T$
dependences of $s_{P_T}$, $s_{v_2}$ and $D_{\epsilon_{2}}$ in
these four hadronaztion situations. } \label{V2_isotrpy}
\end{figure}

In order to investigate the initial geometrical dependence of our
results farther, we next attempt to take Cooper-Frye hadronization
method but with three different anisotropic conditions to see the
change of effect from final hadrons' rescattering. Since the
status of momentum and coordinate space before the hadrons'
rescattering can not  be measured by us, we assume anther three
conditions here.

(I) We decrease the correlation between momentum space and
coordinate space, i.e. increase $\sigma$= 0.2$\pi$ to $\sigma$ =
0.5$\pi$. As we see the circles and squares in figure 5, the
stronger correlation seems to produce more hadrons and suppress
$v_2$ more strongly, but its change for the anisotropy of
coordinate space $\epsilon_{2}$ is slight.

(II) We only remain anisotropy of coordinate space when
hadronization takes place. What will happen after rescattering
among hadrons? The diamons in figure 5 show our results.
Obviously, only hadronic rescatterring  can not induce $v_2$ if
the interaction system has not anisotropy of momentum space before
the hadronic rescattering, though its high $s_{P_T}$ indicates
that more secondary hadrons are produced in rescattering. On the
other hand, the anisotropy of coordinate space also disappear
after hadrons' rescattering. It is also reasonable because the
rescattering plays a role that it decreases $v_2$ and tends to
make the reaction system isotropic from above, which also is
consistent with the isotropic feature of rescattering angle in
LUCIAE rescattering model. So the anisotropy of momentum space,
i.e. $v_2$, may be produced in partonic state before hadronization
if the hadronic rescattering has no direction-sense. In other
words, flow must be formed before the hadronization.

(III) We only remain initial anisotropy of momentum space but lack
initial  anisotropy of coordinate space when hadronization takes
place as shown with triangles in figure 5, the rescattering still
decrease hadrons' $v_2$ and creat a big anisotropy of coordinate
space $\epsilon_{2}$.

Not only in above three conditions but also in our default
condition, we all find that hadrons' rescattering produces many
secondary hadrons and suppresses the elliptic flow $v_2$ of
hadrons to a certain extent. When the two anisotropy of coordinate
and momentum space tend to be homologic with time, the
rescattering ends and the reaction system freezes out.

However, we note that the effect of rescattering on $v_2$ in high
$p_T$ domain is not perfect from Figure 3, i.e., $v_2$ does not
overlap with experimental data which saturates above $P_T \sim$2
GeV/c. We think that one cause may be from our anisotropic factor
in the momentum space. Even though the experimental data can be
fitted well in lower $p_T$, it is a bit artificial since the
nucleon distribution of transverse plane (x,y) and anisotropic
hydrodynamical evolution are not taken into account in this
hydrodynamics \cite{Kolb2}. The other cause could be the partonic
effect on flow, such as partons cascade \cite{Lin} and jet
quenching \cite{Gyulassy,Wang}, is also very important for the
character of particles with high $p_T$, which we also do not take
into account. Of course, the main goal of this work is to
investigate the hadronic rescattering effect on the final state
hadronic elliptic flow.

\section{ Conclusions}

In conclusion, we apply hydrodynamics model followed by the
rescattering model to investigate the effect of hadronic
rescattering on elliptic flow $v_2$ for 200 GeV/c Au + Au at
20-40$\%$ centrality. Hadronization is treated by Cooper-Frye
method and an initial anisotropic phase space of hadrons is
assumed. We find that the yield of hadrons is increased after the
rescattering. While, the rescattering among the hadrons plays a
suppression role  for $v_2$, which makes an asymmetric system in
momentum space less anisotropic. The suppression becomes weaker
with the increasing of transverse momentum. We miss at least 20
$\sim$ 40\% $v_2$ after hadronic rescattering in our hydrodynamics
+ rescattering model. This may, however, change if hadronization
occurs when partons make their last scattering as treated in Texas
multi-phase transport model when the spatial anisotropic  is small
\cite{SAMPT,Chen} instead of when the energy reaches certain
critical energy density presented here. On the other hand, the
hadronic rescattering makes the coordinate space of hadrons tend
to the similar directions to the momentum space. In our model, we
also find that the rescattering does not induce the elliptic flow
if only initial anisotropy of  coordinate space exists in the
hadronic phase.

Considering that the hadronic system created in RHIC collision is
very dense, the following hadronic rescattering may happen
potentially and play an important role. Of course, only
information of the momentum space after the freeze-out can be
measured in experiment. But anyway, the effect of elliptic flow
$v_2$ from final hadronic rescattering deserve attention if the
hadronization takes place when the energy reaches certain critical
energy density presented here in hydrodynamics.

\section*{Acknowledgements}We would like to thank Dr. X. N.
Wang and H. Z. Huang for discussions in earlier stage of this
work. This work was supported partially by the National Natural
Science Foundation of China under Grant No 10328259 and 10135030,
the Chinese Academy of Sciences Grant for the Distinguished Young
Scholars of the National Natural Science Foundation of China under
Grant No 19725521 and the Major State Basic Research Development
Program of China under Contract No G200077404.

{}

\end{document}